\pdfoutput=1
\documentclass[conference]{IEEEtran}
\usepackage[utf8]{inputenc} 
\usepackage{cite}
\usepackage[pdftex]{graphicx}
\usepackage{algorithm}
\usepackage{algorithmic}
\usepackage[tight,footnotesize]{subfigure}
\usepackage[]{caption}
\usepackage[font=footnotesize]{subfig}
\usepackage{epstopdf}
\usepackage{url}
\usepackage{verbatim}
\usepackage{array}
\usepackage[para]{footmisc}
\begin{document}
\title{Studying User Footprints in \\ Different Online Social Networks}
\author{
\IEEEauthorblockN{Anshu Malhotra$^\dagger$ \hspace{1 mm} Luam Totti$^\star$ \hspace{1 mm} Wagner Meira Jr.$^\star$ \hspace{1 mm} Ponnurangam Kumaraguru$^\dagger$ \hspace{1 mm} Virgílio Almeida$^\star$ \hspace{1 mm}}
\IEEEauthorblockA{\\$^\dagger$ Indraprastha Institute of Information Technology \hspace{10 mm} $^\star$Departamento de Ciência da Computação\\
\hspace{20 mm} New Delhi, India \hspace{40 mm} Universidade Federal de Minas Gerais \\
\hspace{10 mm} \{anshu1001,pk\}@iiitd.ac.in \hspace{45 mm} Belo Horizonte, Brazil \hspace{10 mm}\\
\hspace{85 mm} \{luamct,meira,virgilio\}@dcc.ufmg.br}
}
%
\maketitle
\begin{abstract}
With the growing popularity and usage of online social media services, people now have accounts (some times several) on multiple and diverse services like Facebook, LinkedIn, Twitter and YouTube. Publicly available information can be used to create a digital footprint of any user using these social media services. Generating such digital footprints can be very useful for personalization, profile management, detecting malicious behavior of users. A very important application of analyzing users' online digital footprints is to protect users from potential privacy and security risks arising from the huge publicly available user information. We extracted information about user identities on different social networks through Social Graph API, FriendFeed, and Profilactic; we collated our own dataset to create the digital footprints of the users. We used username, display name, description, location, profile image, and number of connections to generate the digital footprints of the user. We applied context specific techniques (e.g. Jaro Winkler similarity, Wordnet based ontologies) to measure the similarity of the user profiles on different social networks. We specifically focused on Twitter and LinkedIn. In this paper, we present the analysis and results from applying automated classifiers for disambiguating profiles belonging to the same user from different social networks. 
 UserID and Name were found to be the most discriminative features for disambiguating user profiles. Using the most promising set of features and similarity metrics, we achieved accuracy, precision and recall of 98\%, 99\%, and 96\%, respectively. 
\end{abstract}
%
\IEEEpeerreviewmaketitle
\section{Introduction}
%
%




In this digital age, when we all are living two lives,  offline and online, our paper trails and digital trails coexist. Our digital trail captures our interactions and behaviors in the digital environment. 
In this virtual age, an important component of our footprints are our online digital footprints~\cite{irani}. Due to the tremendous growth in online social media, the size of a users' online digital footprints is also growing. We interact with friends, post updates about our day to day lives, bookmark links, write online blogs and micro blogs, share pictures, watch and upload videos, read news articles, make professional connections, listen to music, tag online resources, share location updates and what not. Our online digital footprints capture our online identity and whatever we do on the web becomes part of our online identity forever. 



Due to strong relationship between users' offline and online identity~\cite{zafarani,rowe2}, users' online digital footprints can help in uniquely identifying them. By uniquely identifying users across social networks we can discover and link her multiple online profiles. Linking together users' multiple online identities has many benefits e.g. profile management~\cite{carmag1} like managing setting \footnote{\url{http://blisscontrol.com/}} and building a global social networking profile,~\footnote{\url{http://www.digfoot.com/}} help user monitor and control her personal information leakage \cite{rowe2}, user profile portability \cite{szomszor}, personalization \cite{golbeck,carmag1,rowe,szom}. In addition,  linking users' multiple online profiles facilitates analysis across different social networks \cite{golbeck} which helps in detecting and protecting users from various privacy and security threats arising due to vast amount of publicly available user information. 


Unification of users' multiple online identities can lead to various privacy and security threats like: identity thefts and profile cloning \cite{rowe2,bilge} which can lead to compromised accounts \cite{irani}; directed spam and phishing \cite{balduzzi}; online profiling by advertisers and attackers \cite{perito}; online stalking \cite{irani}. An interesting attack was demonstrated by PleaseRobMe.com~\footnote{\url{http://pleaserobme.com/}} where they used Tweets containing Check-Ins from FourSquare to discover if the user was away from her home.
We believe, formulating users' digital footprints and linking her multiple online identities can help to keep the user informed about such threats and suggest her preventive measures. 

The users of the social networks can choose the usernames they wish to, which may be totally unrelated to their real identity \cite{tereza}, and also users may choose different (and unrelated) usernames on different services. People with common names tend to have similar usernames \cite{zafarani, perito}. Users may enter inconsistent and misleading information across their profiles \cite{tereza}, unintentionally or often deliberately in order to disguise. Each social network has different purpose and functionality.  
Heterogeneity in the network structure and profile fields between the services becomes a complicating factor in the task of linking online accounts. All these factors make user profile linking across social networks a challenging task in comparison to Named Entity Recognition \cite{kalashnikov}. 

In this work, we propose a scalable and automated technique for disambiguating user profiles by extracting his online digital footprints from publicly available profile information. The major contributions of our work are:
\begin{itemize}
\item We propose the use of automated classifiers to classify the input profiles as belonging to the same user or not. 
\item Our approach works on publicly available data and does not require user authentication or standardization by different social networks. Sophisticated similarity metrics were used to compare different categories of profile fields. 


\item We conduct a large scale analysis of our approach for linking user accounts across Twitter and LinkedIn, the second and the third most popular social networking sites.~\footnote{\url{http://www.ebizmba.com/articles/social-networking-websites}} We also evaluate our systems' performance in real world.


\end{itemize}

In the next section, we discuss the closely related work. Section~\ref{section:userdisambiguation} describes our system for user profile disambiguation. In section~\ref{section:dataset}, we provide details on the dataset construction. Section~\ref{section:onlinedigitalfootprints} discuss the user profile features and the corresponding similarity metrics used to compare them. Results and analysis from our experimental evaluation is presented in Section~\ref{section:evaluation}. Section~\ref{section:discussion} discusses the conclusions and main findings of our work.

\section{Related Work}\label{section:relatedwork}
Different techniques have been proposed for user disambiguation across social networks. In this section, we discuss the techniques, their limitations and compare them with our approach. Various graph based techniques have been suggested for unifying accounts belonging to the same user across social networks \cite{golbeck, rowe, rowe1, rowe2}. Golbeck et al. generated Friend Of A Friend (FOAF) ontology based graphs from FOAF files / data obtained from different social networks \cite{golbeck} and linked multiple user accounts based on the identifiers like Email ID, Instant Messenger ID. The majority of the analysis was done for blogging websites. Rowe et al. applied graph based similarity metrics to compare user graphs generated from the FOAF files corresponding to the user accounts on different social networks; if the graphs qualified a threshold similarity score, they were considered to be belonging to the same user~\cite{rowe}. They applied this approach to identify web references~/~resources belonging to users \cite{rowe1,rowe2}. Such FOAF graph based techniques might not be scalable and FOAF based data might not be available publicly for all social networks and all users. 

Researchers have also used tags created by users (on different social networking sites such as Flickr, Delicious, StumbleUpon) to connect accounts using semantic analysis of the tags\cite{szomszor, szom, tereza}. While using tags, accuracy has been around 60 -- 80 \%. Zafarani et al. mathematically modeled the user identification problem and used web searches based on usernames for correlating accounts \cite{zafarani} with an accuracy of 66\%. Another probabilistic model was proposed by Perito et al. \cite{perito}. User profile attributes were used to identify accounts belonging to the same user \cite{zafarani, carmag, carmag1, carmag2, vosecky, kontaxis, irani}. Carmagnola et al. proposed a user account identification algorithm which computes a weighted score by comparing different user profile attributes, and if the score is above a threshold, they are deemed to be matched.
 Vosecky et al. proposed a similar threshold based approach for comparing profile attributes \cite{vosecky}. They used exact, partial and fuzzy string matching to compare attributes of user profiles from Facebook and StudiVZ and achieved 83\% accuracy. Kontaxis et al. used profile fields to detect user profile cloning \cite{kontaxis}. They used string matching to discover exactly matching profile attributes extracted by HTML parsing. Irani et al. did some preliminary work
to match publicly available profile attributes to assess users' online social footprints~\cite{irani}. However, their work was limited to categorical and single value text fields and did not include free text profile fields, location and image. Even they suggested as their future work, the use of sophisticated matching techniques and a flexible similarity score for matching profile fields instead of a binary match or a non match decision.

Some of the major limitations of the techniques discussed above are -- specificity to certain type of social networks, e.g., blogging website, networks which support tagging; dependency on identifiers like email IDs, Instant Messenger IDs which might not be publicly available for most of the networks; computationally expensive; use of simple text matching algorithms for comparing different types of profile fields. Also, assignment of weights and thresholds is very subjective and it might not be scalable with the growing size and number of social networks. Most importantly, all the above techniques have been tested on a small dataset (the biggest being 5,000 users), wherein the data collection and evaluation was done manually in some approaches. Lastly, with growing privacy awareness, most of the fields (like gender, marital status, date of birth) which have been used in most of the techniques above, might not be publicly available at present or in future. Real world evaluation has not been done for most of the above techniques. In our work, we address these concerns and hence improve the process of user disambiguation across networks.

\section{User Profile Disambiguation}\label{section:userdisambiguation}
\begin{figure}
\includegraphics[scale=0.21]{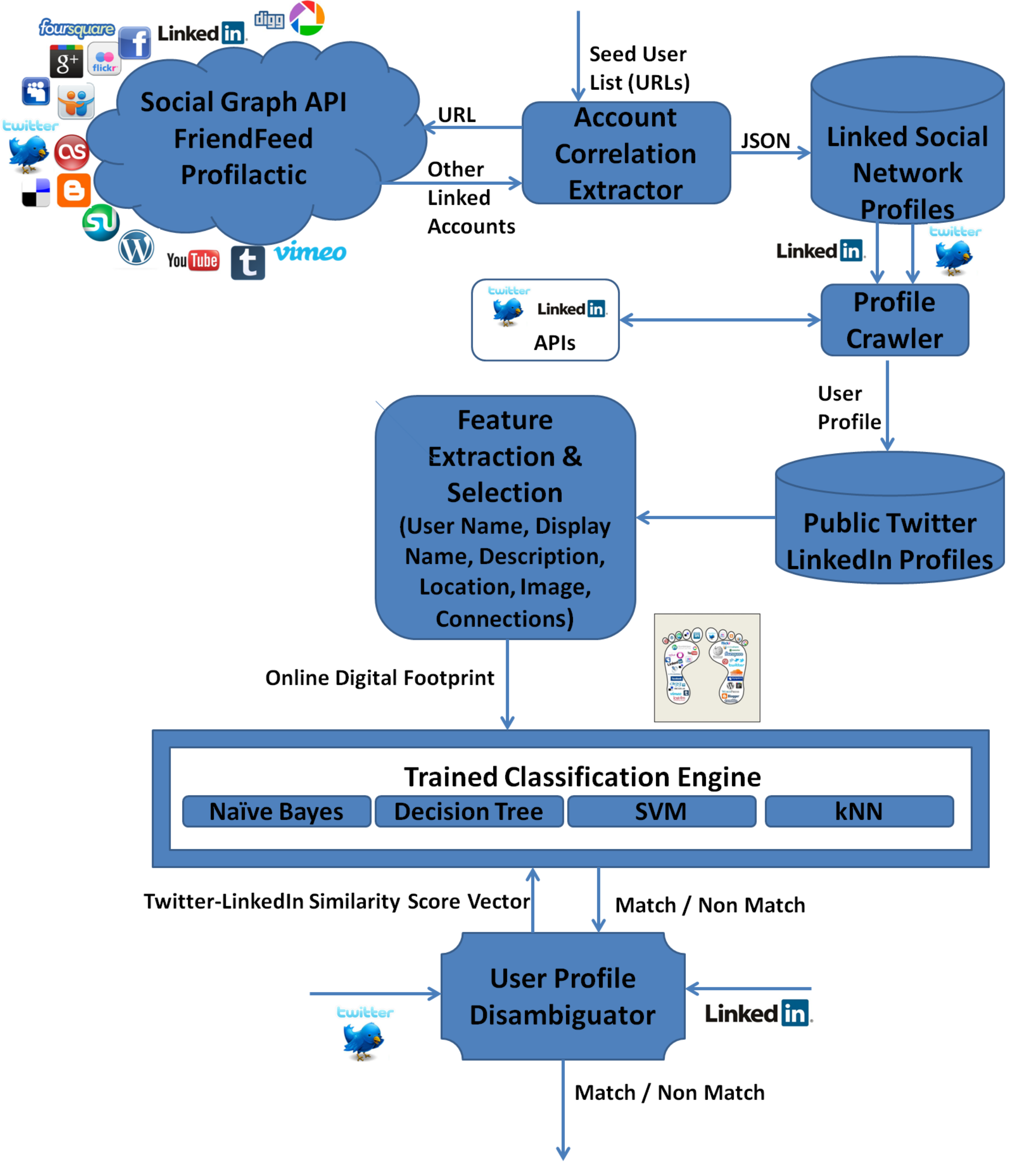} 
\caption{\small{System Architecture. Account Correlation Extractor and Profile Crawler helped in dataset collection. Features were extracted and the Classification Engine was trained using selected features. User Profile Disambiguator was used for system evaluation.}}
\label{system}
\end{figure}

User's digital footprints within a service is the set of all (personal) information related to her, which was either provided by the user directly or extracted by observing the user's interaction with the service. In this paper, we investigate how to match a users' digital footprints across 
different online services, aggregate online accounts belonging to the user across different services, and hence assemble a unified and hopefully richer online digital footprints of the users.
For the purpose of automating this task, we employed simple supervised techniques. Using a 
dataset of paired accounts known to belong to a same user, we compared corresponding features 
from each social network using feature-specific similarity techniques. Each pair of accounts belonging
to a same entity generated a similarity vector in the form
$<username_{score}, name_{score}, description_{score}, \\ location_{score}, image_{score}, connections_{score}>$, where $f_{score}$ is the similarity score between the field \emph{f} (e.g. location) of the user profile in both services. This vector was used as a training instance for supervised classifiers. Similar vectors were generated for profile pairs known to belong to different users. We test the use of these vectors with four classifiers: Naïve Bayes, kNN, Decision Tree and SVM. Our system architecture is depicted in the Figure \ref{system}. Account Correlation Extractor collates the user profiles known to be belonging to the same user across different social networks. Profile Crawler crawls the public profile information from Twitter and LinkedIn APIs for paired user accounts for these services. A user's Online Digital Footprints are generated after Feature Extraction and Selection. Various classifiers are trained for account pairs belonging to same users and pairs belonging to different users, which are then used to disambiguate user profiles i.e. classify the given input profile pairs to be belonging to the same user or not. 

\section{Dataset}\label{section:dataset}
In this section, we discuss the dataset we collated for developing automated mechanisms for disambiguating digital footprints of a user. 
\subsection{Collection}
The data collection consisted of two phases. The first stage involved collecting the true positive connections, i.e. the profiles from different services known to belong to the same user. We used the following sources in this phase:
\paragraph*{Social Graph API}\footnote{http://code.google.com/apis/socialgraph/docs/} This Google API provides
access to declared connections between public URLs. A URL can be a website or a user's profile page. When connections for a given user profile URL are requested, Social Graph returns other profile URLs that 
are alternative identities of the requested user, allowing us to retrieve accounts of a same user across multiple services. We collected information of around 14 million users, although only 28\% of those users
had useful declared connections.
\paragraph*{Social Aggregators} Social network aggregators are services that pull together the 
feeds from multiple social networks that the user manually configured. We crawled 883,668 users 
from FriendFeed~\footnote{http://friendfeed.com/} and 38,755 users from 
Profilactic.~\footnote{\url{http://www.profilactic.com/}}


After this first stage of data collection, we had true positive connections of the profiles belonging to 
the same user across different social networks. This data for each user $U$ was a N-tuple record of the 
form $(u_1^{s_1}, u_2^{s_2}, \dots, u_n^{s_n})$, where $s_i$ is the online service in which user
$U$ has an account using the identifier $u_i$.
We wanted to measure the effectiveness of different features of a user profile in forming unique and distinguishable digital footprints of a user. For this, we formulated a model to disambiguate a user on a social network given his digital from some other social network. To accomplish these tasks, we required the profile features of the user profiles from different social networks. This comprised the second phase of our data collection. Using the unique user handle, we crawled and collected the publicly available profile fields of the users.\footnote{All profiles declared as non-English were ignored (around 13\%).}

\subsection{Data Summary}
\begin{figure}[h]
\centering
\includegraphics[scale=0.5]{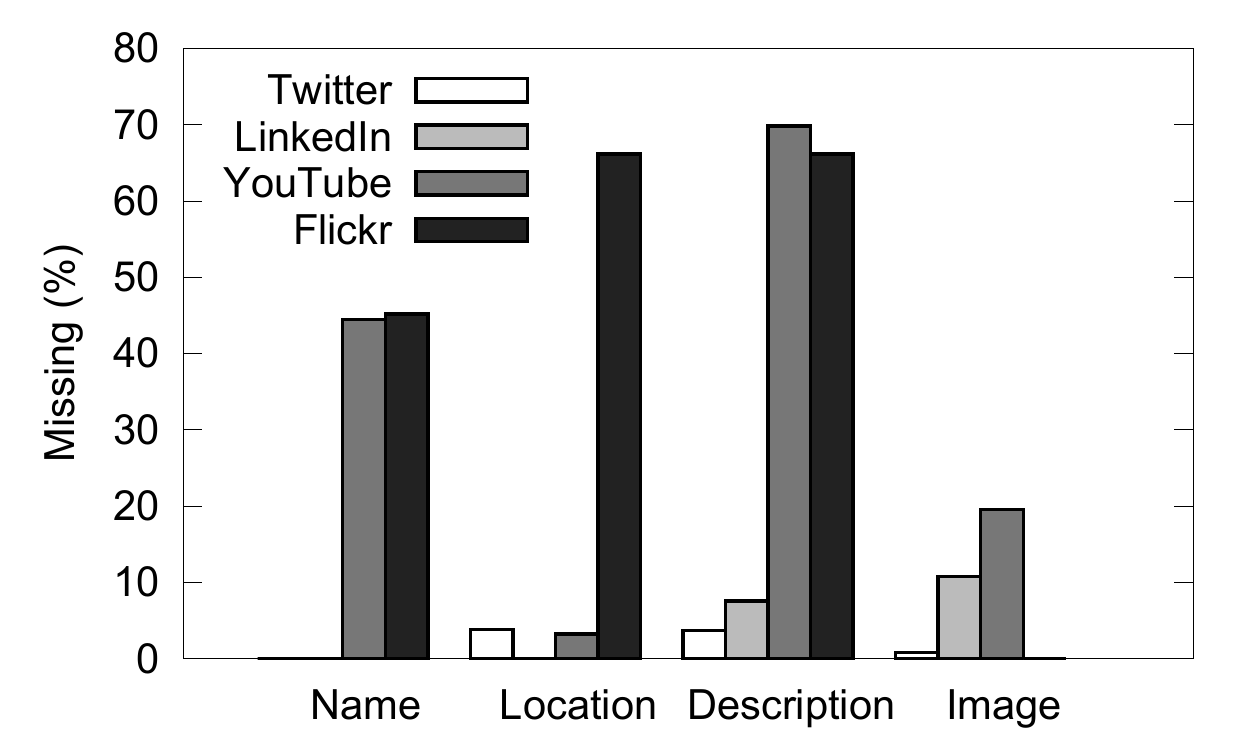}
\caption{\small{Percentage of missing features in each service.}}
\label{fig:missing}
\end{figure}

To start with, our data consisted of 41,336 user profiles from each
of the following services: Twitter, YouTube and Flickr. Each account triple
were know to belong to the same user. Additionally, 29,129 pair of accounts
were collected from Twitter and LinkedIn. 
We observed that profile information from YouTube 
and Flickr had large proportion of 
missing fields (Figure \ref{fig:missing}). We used 29,129
accounts from Twitter and LinkedIn for all further analysis in this work. \\

\section{Online Digital Footprints}\label{section:onlinedigitalfootprints}
A user profile on any social networking website can be seen as an N dimensional vector, where each dimension is a profile field, e.g., username, first name, last name, location, description / about me, relationship and others \cite{vosecky}. A subset of these features (e.g. username, location) from one social network can be used to disambiguate the same user from millions of other users on another social network. 
 We chose to study the Twitter-LinkedIn connections since they had some comparable profile features like location, `about me' / description and others. In this section we discuss the features and techniques we used for matching the users' digital footprints across social networks.
\subsection{UserID}
This refers to the unique username / user ID / handle which identifies a user on the social network and in many cases is used by the person to log in to the service. 
A person may have different usernames across social networks and hence cannot be identified with only her username. We employed a sophisticated string matching method, \emph{Jaro-Winkler~\footnote{\url{http://en.wikipedia.org/wiki/Jaro-Winkler_distance}} distance (jw)}
that is designed for comparing short strings and gives a score in the range [0, 1]. Higher the score, higher the similarity of the two strings. 
\subsection{Display name}
This refers to the first name and / or the last name which the user has entered in his profile. 
Instead of exactly matching display names, we again employed the \emph{Jaro-Winkler distance} for computing the similarity between display names. However, users with the same display name might have similar usernames and hence we need to look at other features which can help identify a user. 
\subsection{Description}
This is the short write up / `bio' / `about me' which the user provides about himself. 
We employed the following three methods to compare description fields from two profiles which were to be matched -- \emph{tf-idf vector space model}: 
The description fields were first pre-processed by removing the punctuations, stop words to extract the tokens which were lemmatized and converted to lower case. The cosine similarity was then computed between the two token sets, considering each to be a document, therefore resulting in a score in the range [0,1] between the two fields. \emph{Jaccard's Similarity}: 
Applying the same pre-processing described in the previous method, the Jaccard’s similarity\footnote{\url{http://infolab.stanford.edu/~ullman/mmds/ch3.pdf}} between the two token sets was taken as the similarity score.
\emph{WorldNet based Ontologies}: 
Wordnet\footnote{\url{http://wordnet.princeton.edu/}} is a common English language lexical database 
which provides ontologies i.e. groupings of synsets based on hypernym-hyponym (is-a-relation) tree. This ontology organized using hypernym tree can be used to explain the similarity or dissimilarity between synsets. We use the Wu-Palmer 
similarity metric \cite{wup} between tokens of the description fields from the two user profiles to be matched \cite{prantik}.
\subsection{Location}
The next profile field we chose for comparison was location (loc). 
For comparing the location field, we extracted the tokens from the location field of both the profiles to be matched by removing the punctuations and converting to lower case. For these tokens, we computed the following metrics -- \emph{Sub Strings Score (substr)}: normalized score of number of tokens from one location field present as a sub-string in the other; \emph{Jaccard's Score}: Jaccard's similarity of tokens from two location fields; \emph{Jaro-Winkler Score}; \emph{Geographic Distance (geo)}: Euclidean Distance between the two locations was found using their latitude and longitudes. Latitude and longitude were found by querying Google Maps GeoCoding API.~\footnote{https://developers.google.com/maps}
\subsection{Profile Image}
A profile image is a thumbnail provided by the user for the purpose of visually representing him. 
The collected user profiles provide the URL in which the image is made available. The images were downloaded and stored locally. Each image was then scaled down to 48 x 48 pixels using cubic spline interpolation and then converted to gray scale by taking the scalar product of the RGB components vector $(r,g,b)$ with the coefficient vector $(0.299, 0.587, 0.114)$. Each image could then be seen as a vector of values from 0 to 255 to which simple functions were applied to quantify their similarity. This feature may be abbreviated as `img' throughout the paper. We used \emph{Mean Square Error}, \emph{Peak Signal-to-Noise Ratio}, and \emph{Levenshtein (ls)} for analyzing the profile image.
\subsection{Number of Connections}
The last feature was derived from the intuition that a user has a similar number of friends (which we generalize as connections) across different services. For Twitter, we considered the number of connections of a user $u$ to be the number of users that $u$ follows. For LinkedIn, a user $v$ is a connection of $u$ if $v$ belongs to the private network of user $u$. The number of connections in different services can assume different ranges, with different meanings. For example, a certain number of connections on LinkedIn can mean that a user is very active and popular, while the same number on Twitter can be much less significant. Taking this into consideration, two different techniques were employed to compare those two values -- \emph{Normalized (norm)}: Each connection value $c$ was normalized to the range $[0..1]$ using the smallest and greatest connection values observed in each service. 
The similarity was then taken as the unsigned difference between the two values. \emph{Class}: \emph{norm} is very vulnerable to outliers, e.g. a single big value would compress all the other values into a very small range, possibly suppressing relevant information. To overcome this, each value was assigned a class denoting how big it was. This was done by organizing all connection values into a sorted vector and then dividing it into $k$ equally sized clusters, where $k$ was the chosen number of classes. Once each value was assigned a class index between $1$ and $k$, the similarity was taken as the unsigned different between those two indexes. We adopted $k=5$ in this work.

\section{Evaluation Experiments}\label{section:evaluation}
In this section, we describe and evaluate the experiments performed using the dataset 
and metrics proposed in Section~\ref{section:onlinedigitalfootprints}. The analysis was done using a dataset of 
account pairs for 29,129 unique users. 
\subsection{Feature Analysis}
\begin{figure*}[ht]
\flushleft
\subfigure{ \includegraphics[scale=0.23]{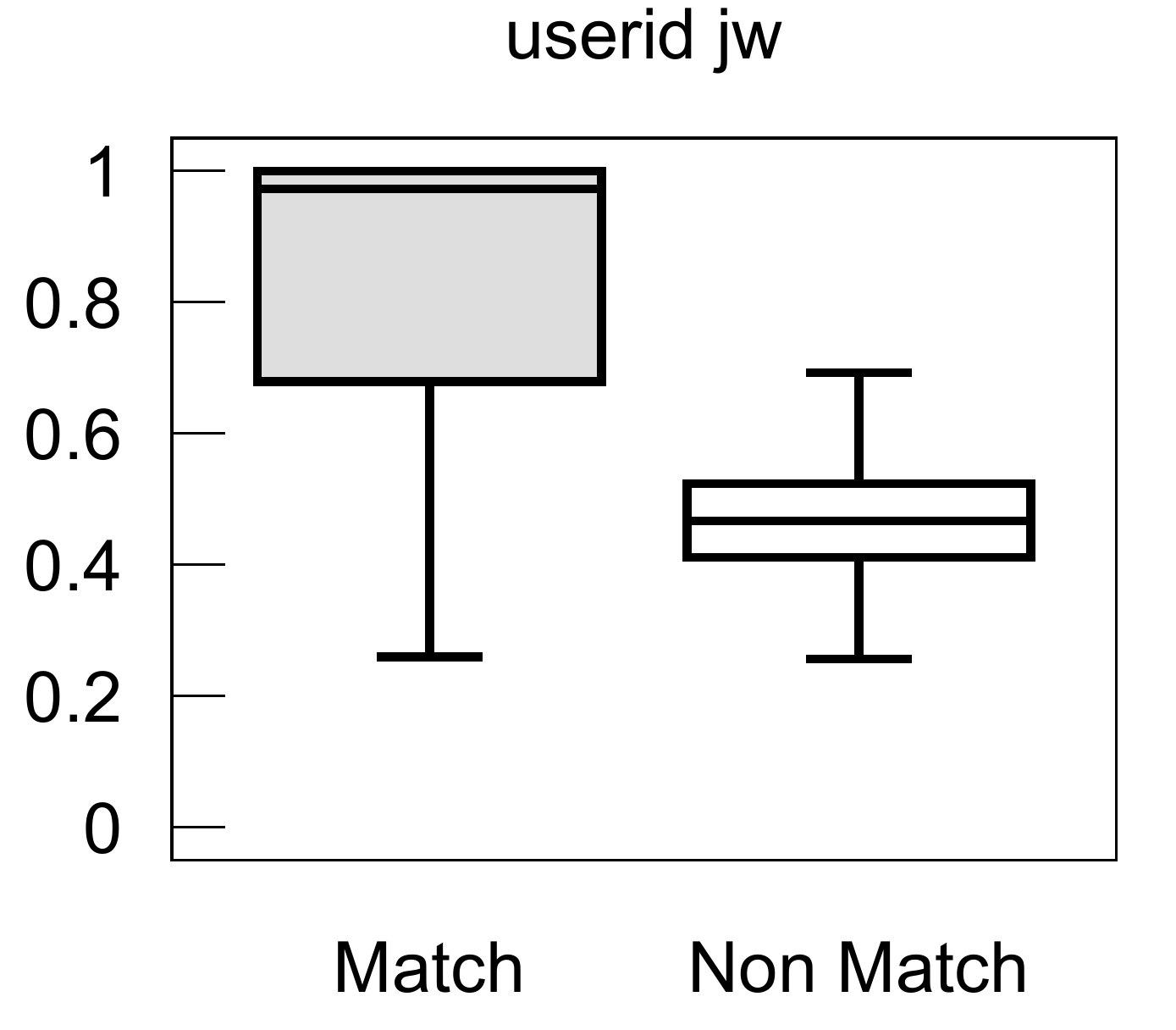} }
\subfigure{ \includegraphics[scale=0.23]{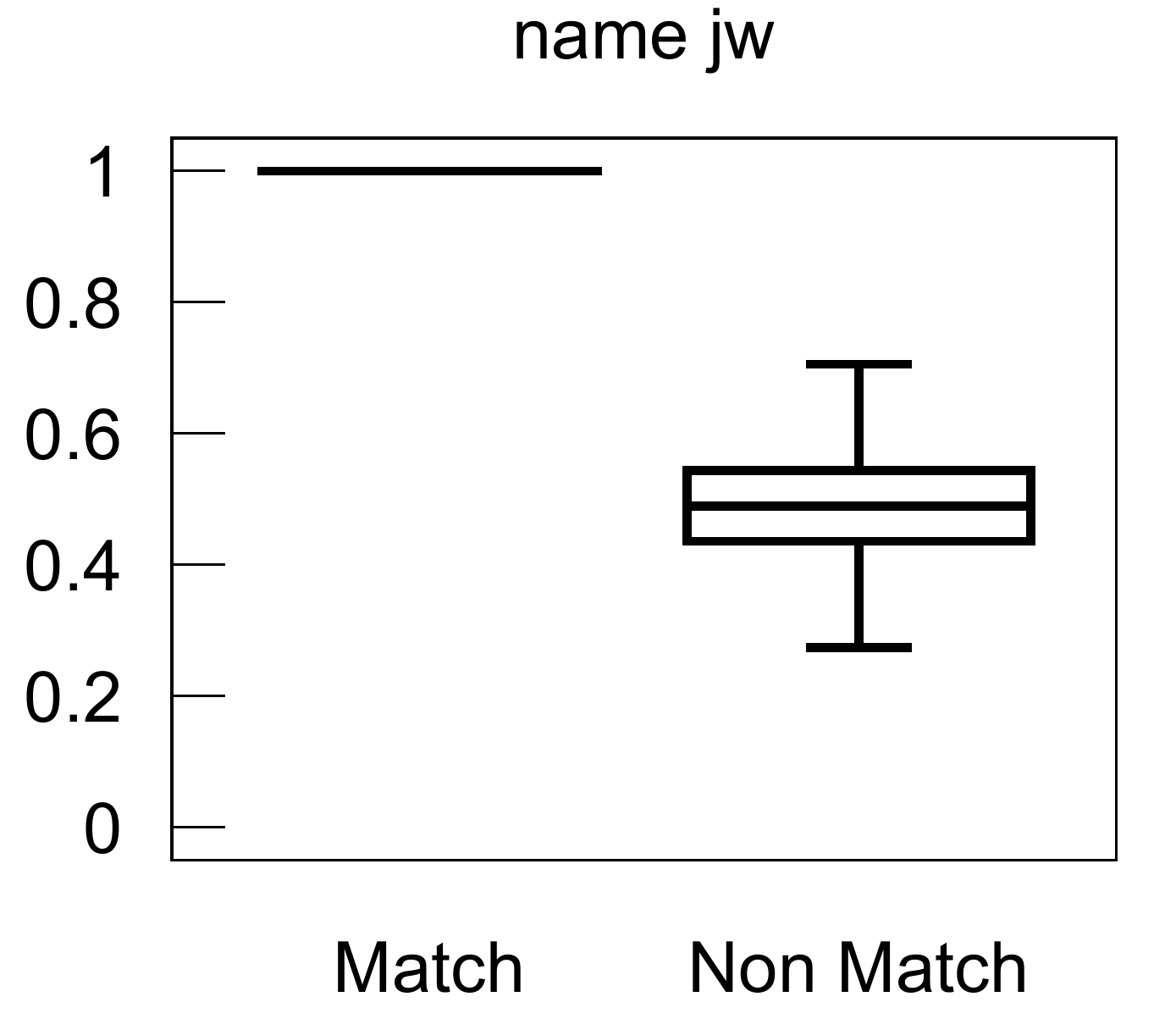} }
\subfigure{ \includegraphics[scale=0.23]{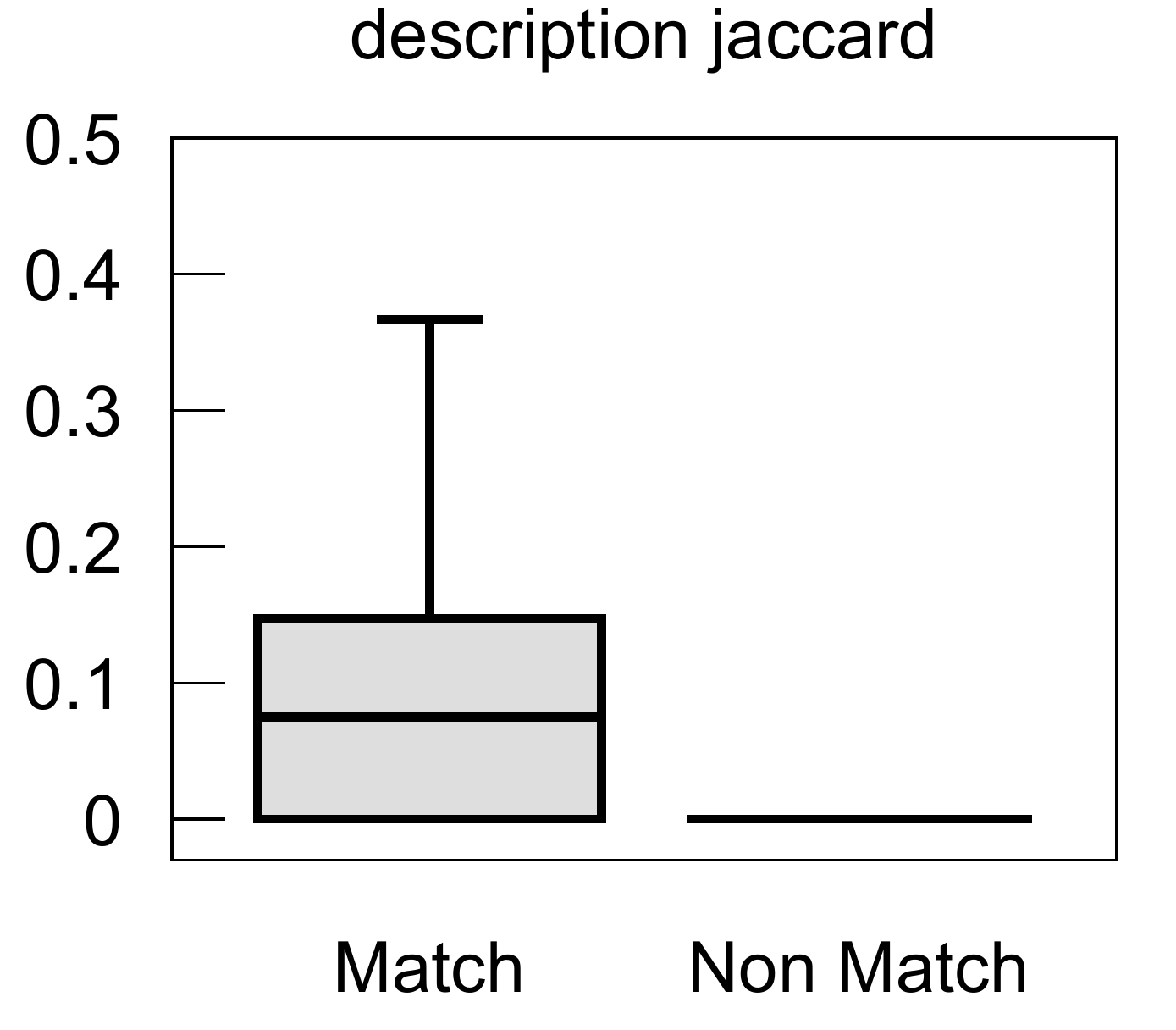} }
\subfigure{ \includegraphics[scale=0.23]{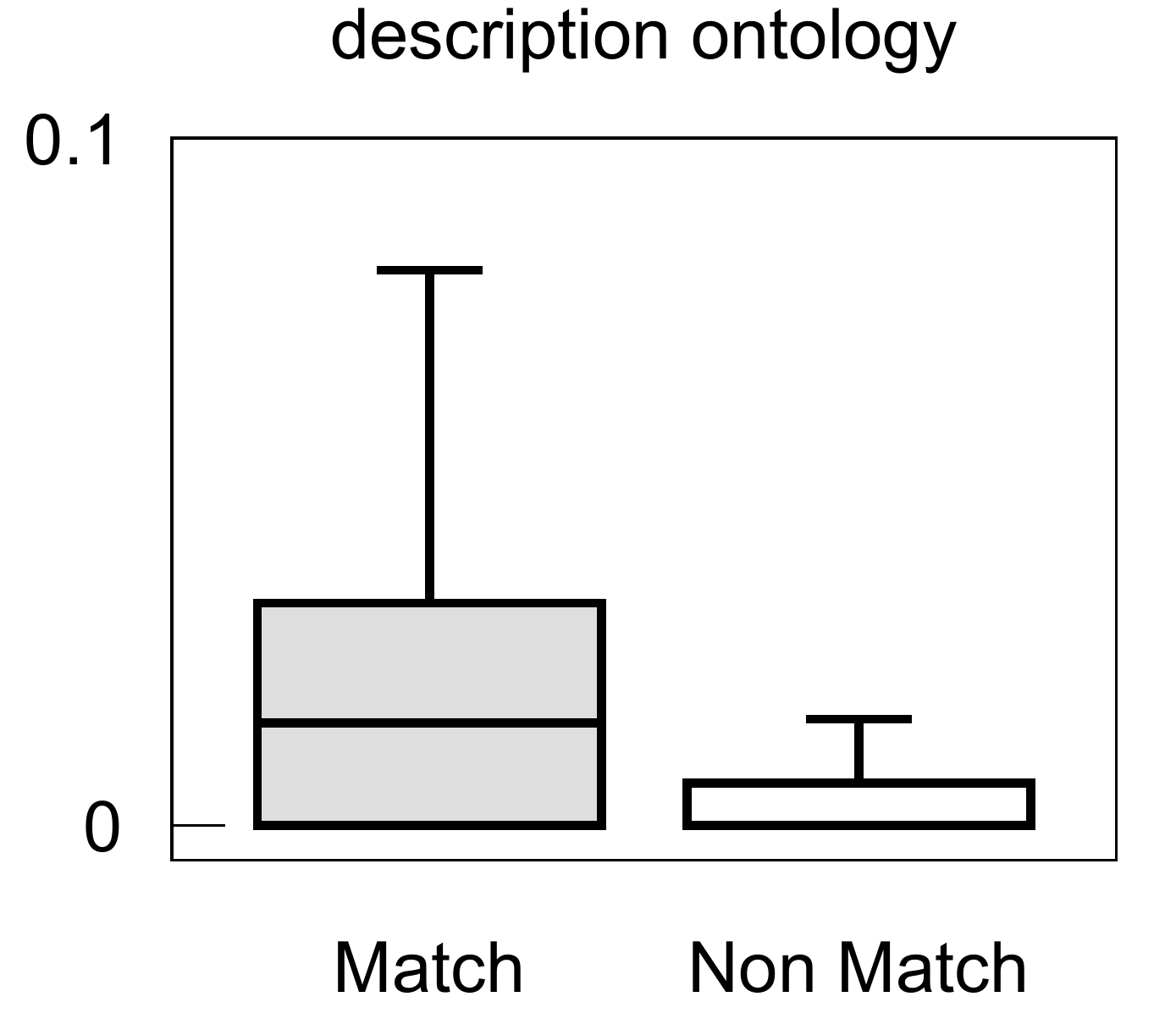} }
\subfigure{ \includegraphics[scale=0.23]{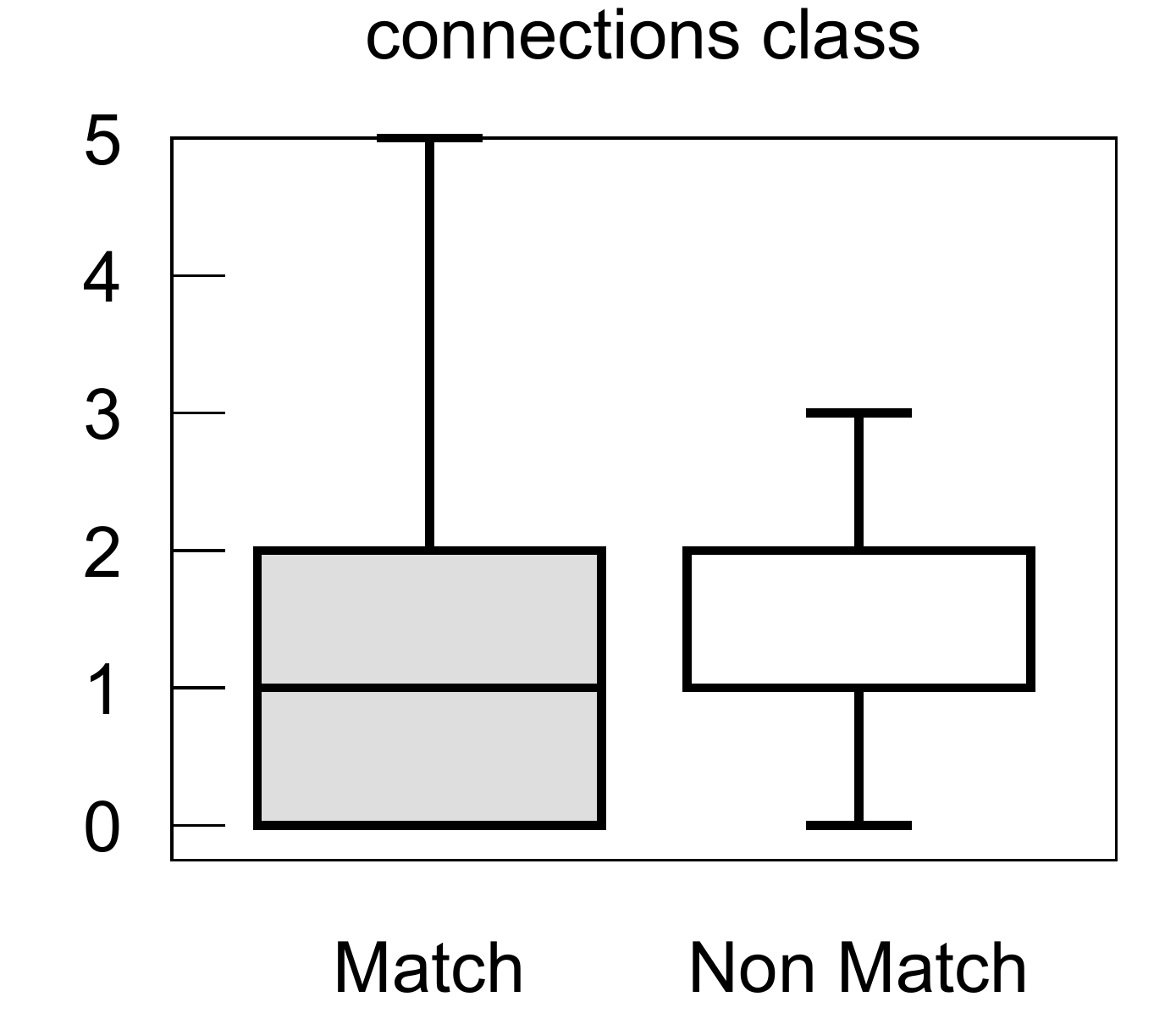} }
\subfigure{ \includegraphics[scale=0.23]{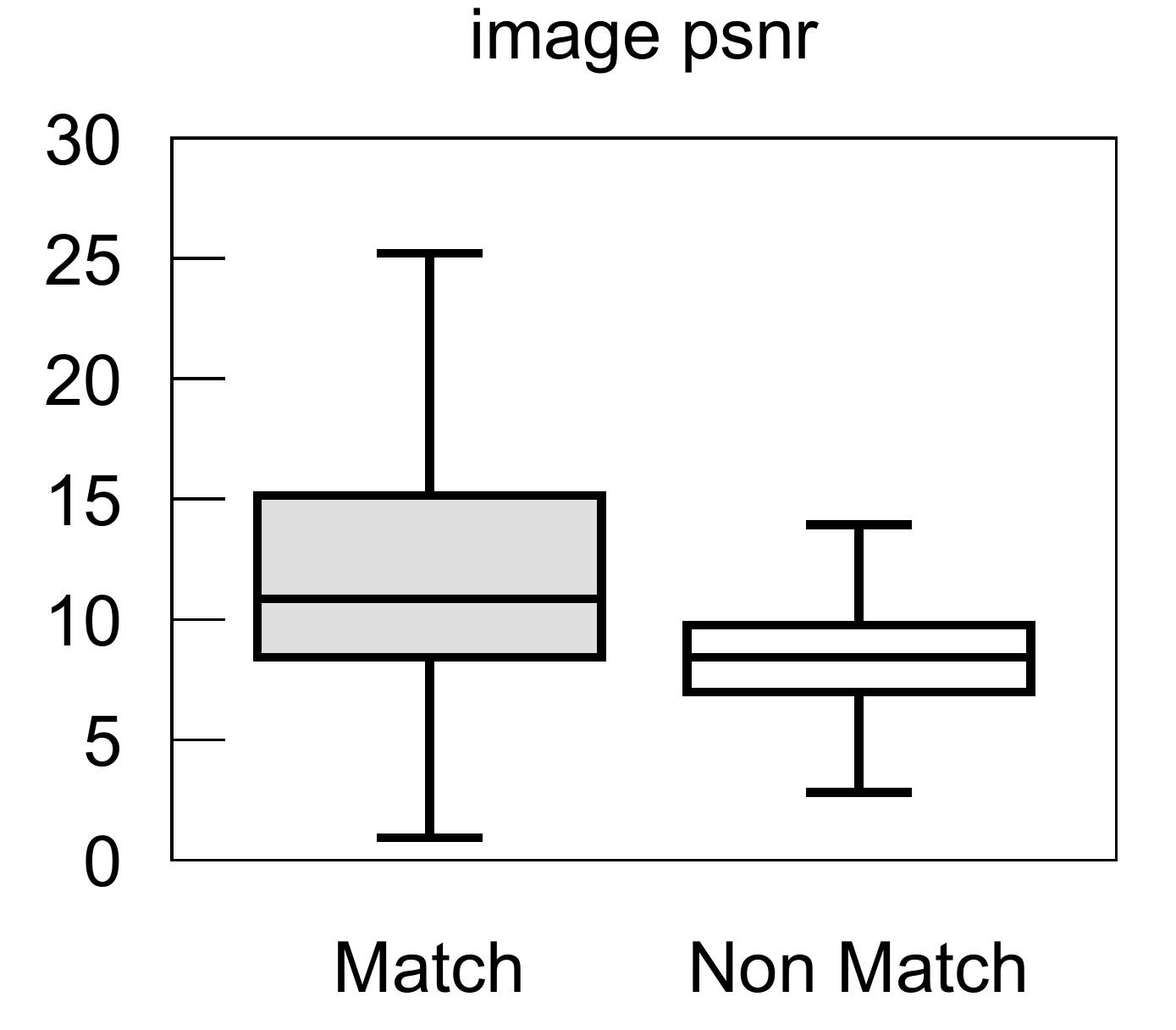} }
\subfigure{ \includegraphics[scale=0.23]{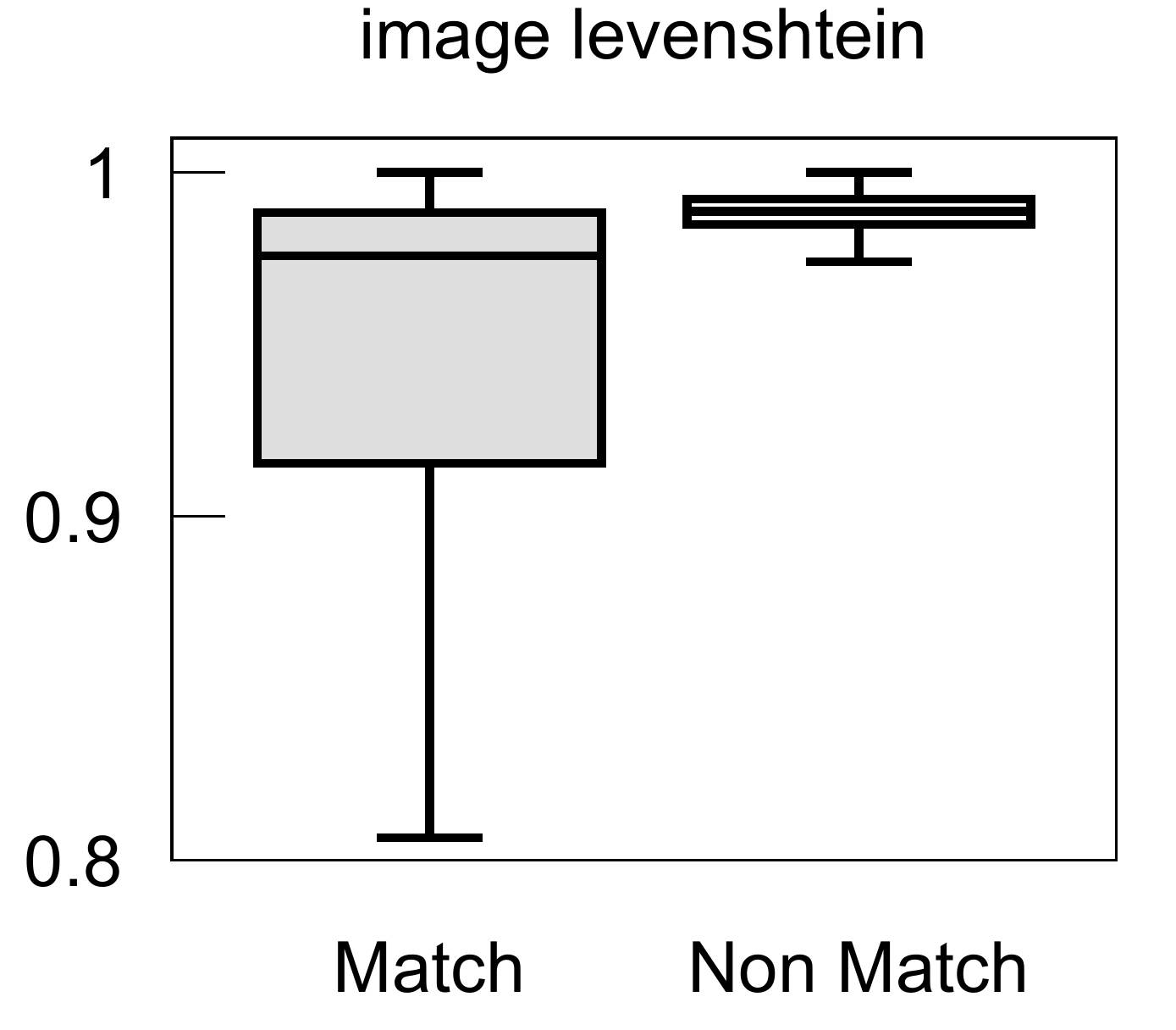} }
\subfigure{ \includegraphics[scale=0.23]{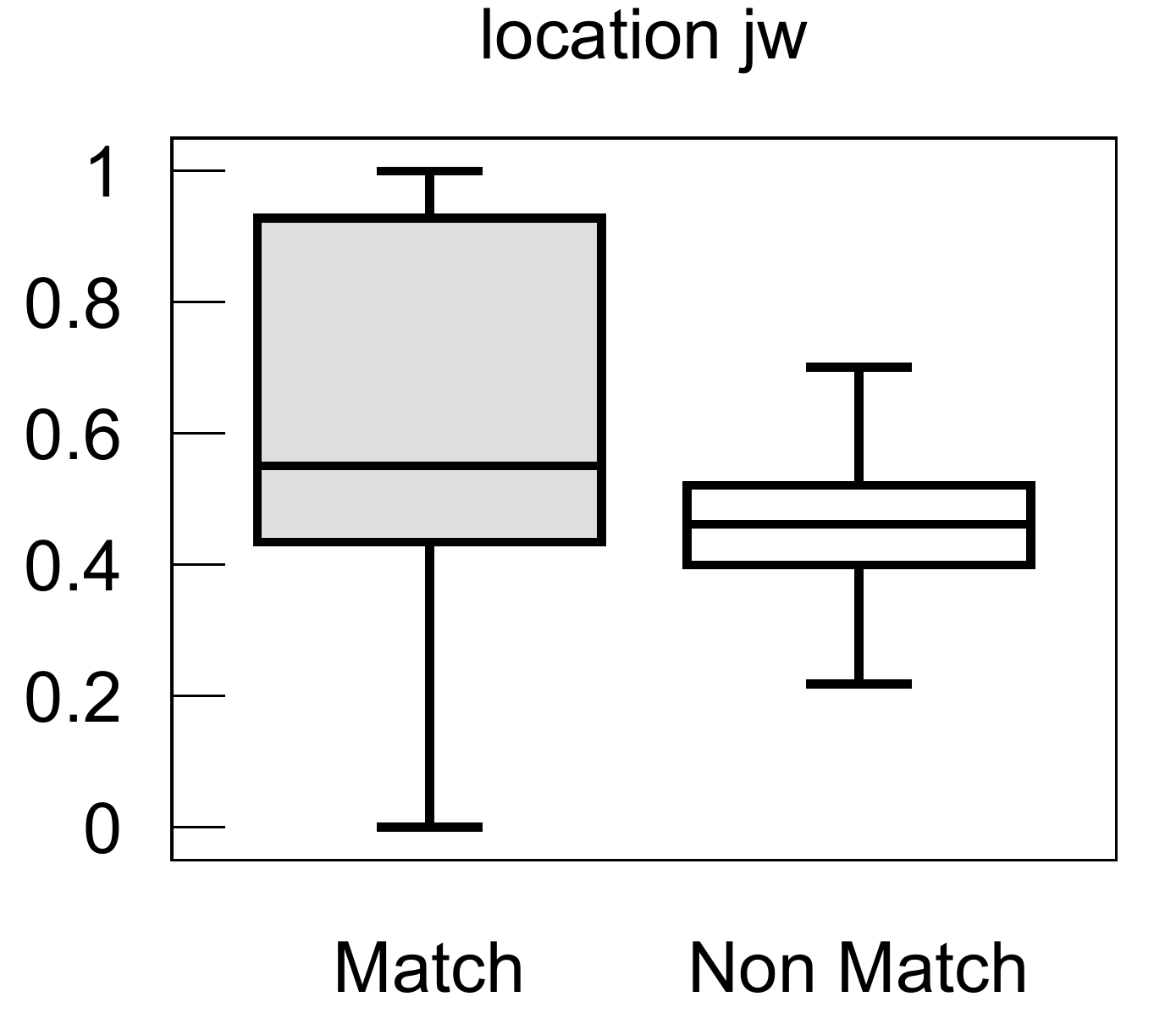} }
\subfigure{ \includegraphics[scale=0.23]{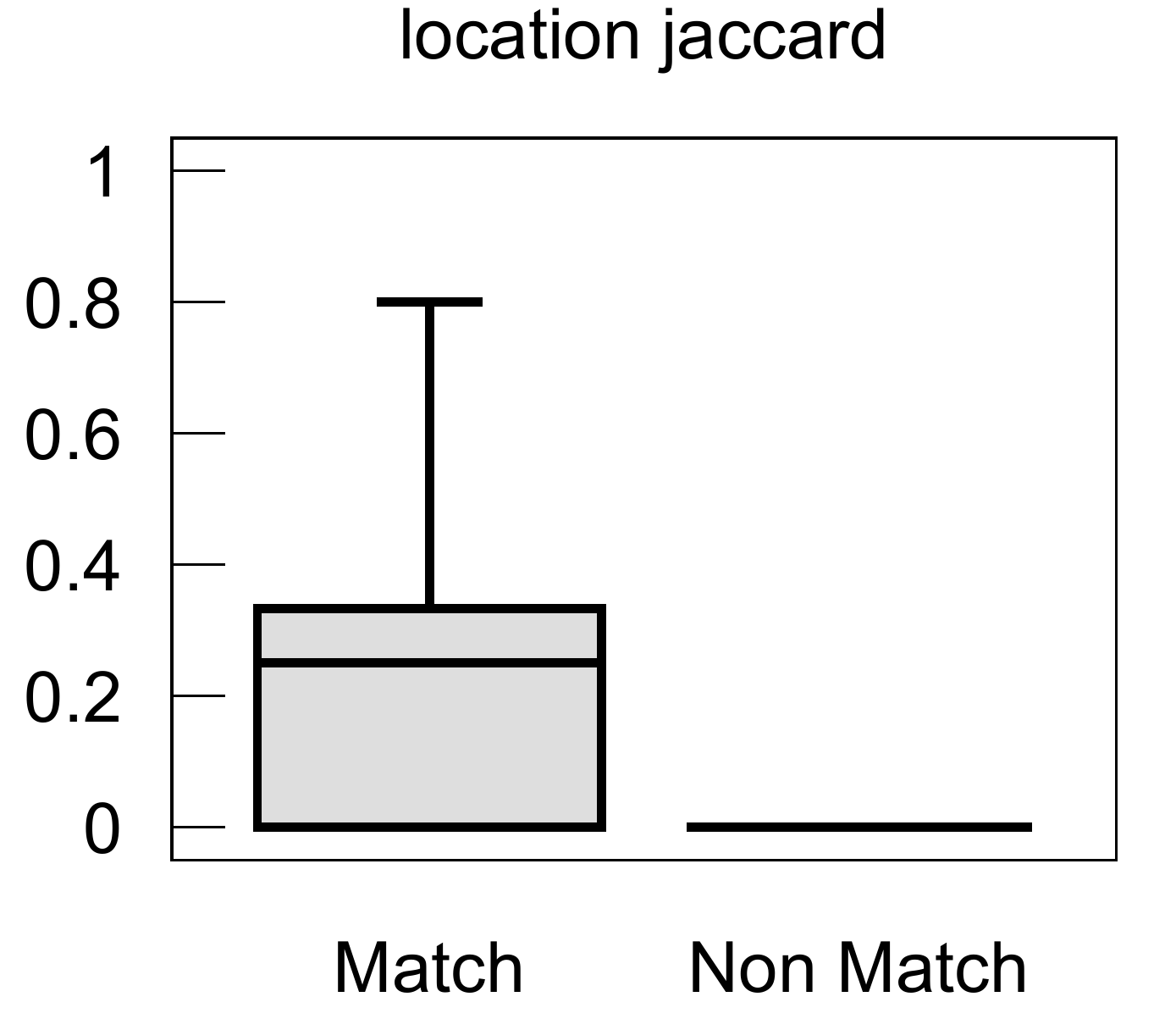} }
\subfigure{ \includegraphics[scale=0.23]{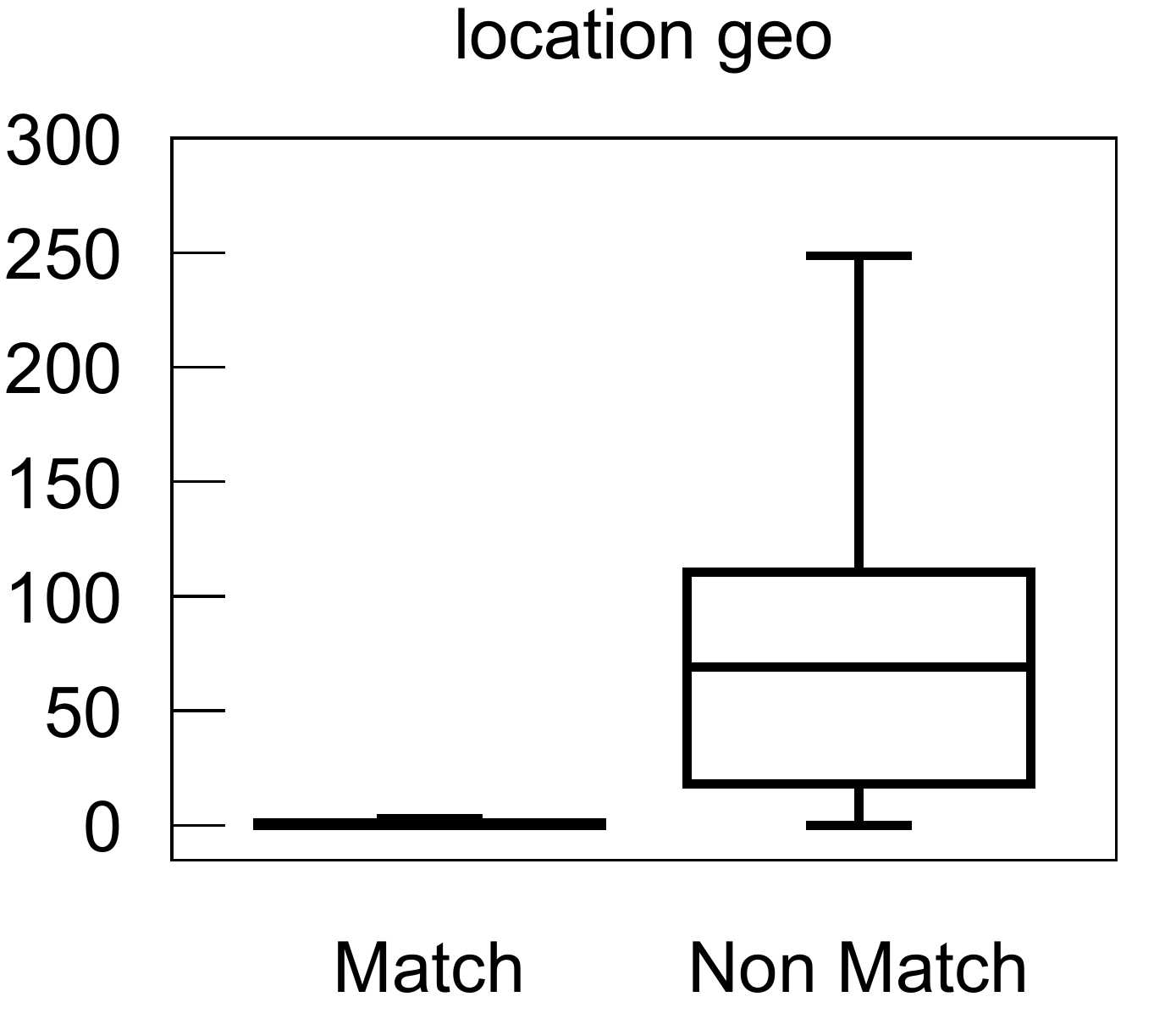} }
\caption{\small{Box plots for each feature separating the values of the ``Match" class and 
the ``Non Match" class.}}
\label{fig:boxes}
\end{figure*}

With the purpose of effectively measuring the similarity between two fields, different approaches were proposed in this work to assess the usefulness of each feature and similarity metric in the classification 
process. Table \ref{tab:features} shows the features' discriminative capacity
according to four different scores: Information Gain (IG), Relief \cite{kira92}, Minimum Description
Length (MDL) \cite{mdl98} and Gini coefficient \cite{gini10}. For the metrics that can only 
be applied to categorical attributes an entropy based discretization approach was used 
\cite{Fayyad_Irani_1993}. Throughout this section, we represent each component of a similarity vector
as the feature name subscripted with the similarity metric used, e.g. $<$userid$_{jw}$, 
desc$_{jaccard}$, loc$_{geo}>$.
For each feature the similarity metric with the highest score is highlighted.
Additionally, box plots are shown in Figure \ref{fig:boxes} for some of the feature
to enlighten how their value distributions affect their discriminative capacity. 
For each feature different boxes are plotted for the values of each class, ``Match" and 
``Non Match." Outliers were omitted for better clarity. In Table \ref{tab:features}, we can 
see a consistency across all four scores in all feature groups. 
\emph{UserID} and \emph{Name} are the most
discriminative features, which are clearly supported by the values distributions on Figure \ref{fig:boxes}.
Both features show very polarized distributions, with no overlap between the ranges of the 
different classes.
For the \emph{Description} values, \emph{tf-idf} has shown to be slightly better than \emph{Jaccard},
while \emph{Ontology} presented poor results. 
The \emph{Geo-Location} metric has shown to be considerably superior than the other metrics 
for the \emph{Location} field. 
For fast and low cost solutions though, \emph{Jaccard} and \emph{Sub-String} 
can be considered viable alternatives. Both implemented metrics for the \emph{Connections} showed low values for all scores, 
which is also supported by the box plots. Manual verification confirmed that the 
intuition that a same user should have a similar number of friends in different social 
networks may be flawed. In particular, for Twitter and LinkedIn this is generally not 
true due to the different nature of the services. 
The \emph{Image} feature presented a small but significant informational relevance, 
being the Levenshtein distance the best metric to be used. 
\begin{table*}[ht]
\begin{tabular}{c!{\vrule width 1.4pt}c!{\vrule width 1.4pt}c!{\vrule width 1.4pt}c|c|c
!{\vrule width 1.4pt}c|c|c|c!{\vrule width 1.4pt}c|c!{\vrule width 1.4pt}c|c|c!
}
\hline
& \textsc{UserID} & \textsc{Name} & 
\multicolumn{3}{c!{\vrule width 1.4pt}}{\textsc{Description}} & 
\multicolumn{4}{c!{\vrule width 1.4pt}}{\textsc{Location}} & 
\multicolumn{2}{c!{\vrule width 1.4pt}}{\textsc{Connections}} & 
\multicolumn{3}{c}{\textsc{Image}}\\ \hline \hline
& JW & JW & Jaccard & \textbf{TF-IDF} & Ontology & JW & Jaccard & 
Substr & \textbf{Geo} & Norm & \textbf{Class} & MSE & PSNR & \textbf{LS} \\ \hline \hline
IG & 0.548 & 0.812 & 0.286 & \textbf{0.323} & 0.161 & 0.232 & 0.337 & 
0.350 & \textbf{0.520} & 0.000 & \textbf{0.009} & 0.183 & {0.184} & \textbf{0.215} \\ 
Relief & 0.434 & 0.521 & 0.134 & \textbf{0.180} & 0.113 & 0.108 & 0.041 & 
0.039 & \textbf{0.227} & 0.002 & \textbf{0.095} & 0.157 & {0.158} & \textbf{0.188} \\ 
MDL & 0.379 & 0.562 & 0.274 & \textbf{0.300} & 0.188 & 0.158 & 0.233 & 
0.270 & \textbf{0.488} & -0.006 & \textbf{0.006} & 0.205 & 0.205 & \textbf{0.227} \\ 
Gini & 0.151 & 0.217 & 0.084 & \textbf{0.092} & 0.051 & 0.067 & 0.098 & 
0.102 & \textbf{0.146} & 0.000 & \textbf{0.003} & 0.051 & 0.051 & \textbf{0.061} \\ 
\end{tabular}
\caption{\small{Discriminative capacity of each pair $<feature,metric>$ according to
four different approaches.}}
\label{tab:features}
\end{table*}

\subsection{Matching profiles}
The similarity methods presented in the previous 
section were applied to the accounts collected from Twitter and LinkedIn to produce 
a training set for the classifiers. The positive examples consisted of all the 
similarity vectors for the account pairs of the Social Graph dataset.
An equal number of negative examples were synthesized by randomly 
pairing accounts that don't belong to the same user and calculating their similarity 
vectors. This yielded a total of $58,258$ training instances.
After training the classifier, they were tested by giving them as input a Twitter-LinkedIn 
profile pair to be classified as a ``Match" or a ``Non Match." A ``Match" means that the two given 
input profiles belong to the same user, while ``Non Match" means they don't. The results shown in this
section were obtained by 10-fold cross-validation on the data.
In order to further evaluate the effectiveness of the adopted similarity metrics,
we generated results for all the possible combinations of the features and metrics. 
The feature set with the best accuracy using Naïve Bayes was $<$name$_{jw}$, userid$_{jw}$, loc$_{geo}$, 
desc$_{jaccard}$, img$_{ls}>$. We also observed that the features
\emph{Name}, \emph{UserID} and \emph{Location} using Geo-Location
were present on all of the top 10 results, confirming that they are relevant features.

Table \ref{tab:results} presents detailed results for the most promising set 
of features according to previous results. The results for each classifier are
very comparable, except for the kNN. Using the most promising set of features and similarity metrics, we achieved accuracy, precision and recall as 98\%, 99\% and 96\% respectively.
\begin{table}[h]
\centering
\small
\begin{tabular}{r|c|c|c|c}
& \small{Accuracy} & \small{Precision} & \small{Recall} & \small{$F_1$} \\ \hline \hline
\small{Naïve Bayes} & \textbf{0.980} & 0.996 & \textbf{0.964} & \textbf{0.980} \\
\small{Decision Tree} & 0.965 & 0.994 & 0.936 & 0.964 \\
\small{SVM} & 0.972 & 0.988 & 0.956 & 0.971 \\
\small{kNN} & 0.898 & \textbf{0.998} & 0.798 & 0.887 
\end{tabular}
\caption{\small{Results for multiple classifiers using the feature set \{name$_{jw}$, 
userid$_{jw}$, loc$_{geo}$, desc$_{jaccard}$, img$_{ls}$, conn$_{norm}$\}.}}
\label{tab:results}
\end{table}
\subsection{Finding Candidate User Profiles}
\begin{figure}[ht]
\centering
\includegraphics[scale=0.4]{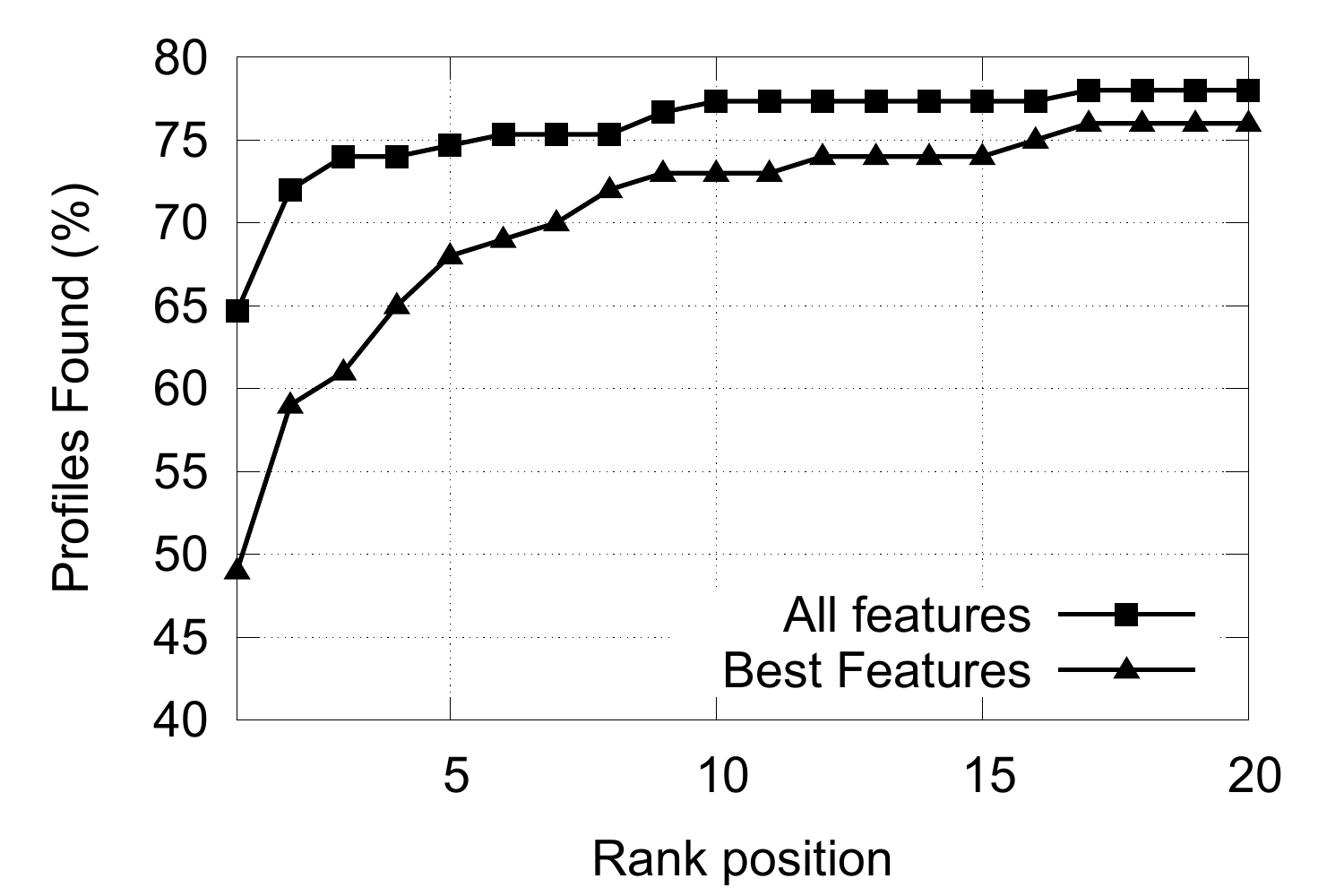}
\caption{\small{Relation between the position in the rank $r$ and the percentage of times the right profile
is found in a position lower or equal to $r$.}}
\label{fig:rank}
\end{figure}

To evaluate how our model for user profile disambiguation would perform in a real scenario,
we developed a system for retrieving account / profile candidates from the services' API,
in order to find a possible match for a known account.
%
More specifically, we reserve a part of the true positive data to be a testing set $T$. 
A classifier $C(v_i)$ is then trained with the remaining dataset. We modified Naïve  Bayes
to return the probability that the similarity vector $v_i$ was generated by 2 profiles
that belong to the same user.
Now, for each instance $<p_{t}, p_{l}> \in T$ we query Twitter's API using the 
LinkedIn display name, $p_{l}[name]$. Let $C$ be all the accounts returned by Twitter.
For each $c_i \in C$ we compute the similarity vector $S(c_i, p_{l}) = v_i$, which is now 
a instance suitable for our model. For each $v_i$, calculate the probability $P_i$ of
$c_i$ belonging to the same user of $p_{l}$, which is basically $C(v_i)$. At last, 
we sort all the values $C(v_i)$ in decreasing order to form a rank $R$ in which, ideally,
$p_{t}$ should be at the top. 
Figure \ref{fig:rank} shows how good our profile ranking mechanism is. The $x$ axis is the
position in the rank and the $y$ axis is the percentage of times the right profile
was found in a position lower or equal to $x$. We plotted curves first using \emph{UserID} and \emph{Name}, and next using all the profile features in order to verify whether using all features was unnecessary. Although the best 
features have shown to be really good discriminators, when doing searches in the services
APIs using a given name, most of the returned accounts have very similar values in the
fields \emph{Name} and \emph{UserID}, making the use of more features specially important.
This assumption was confirmed by the observed results.

Figure \ref{fig:rank} shows that in 64\% of the cases the right profile was found in the first
position of the rank when using all features, while this value was 49\% for the set of the 
best features. 
This shows that our model could be used to match
profiles automatically with a 64\% accuracy rate by choosing the best guess. The system 
could also be used in a
semi-supervised manner to narrow down candidates. For example, we can see that in 75\% 
of the times the right profile was in the first 3 positions of the rank.

\section{Discussion}\label{section:discussion}
In this work, we applied automated techniques along with users' online digital footprints from one social network to identify her on another social network. We extracted the users' online digital footprints entirely from the public profile information. We used multiple profile features and sophisticated similarity metrics to compare them and assessed their discriminative capacity for user profile disambiguation. UserID and Name when compared using the Jaro-Winkler metric were the most discriminative ones. Using the most promising set of features and similarity metrics, we achieved accuracy, precision and recall as 98\%, 99\% and 96\% respectively. We tested our system in real world to find candidate user profiles on Twitter, using the display name of a LinkedIn user. Seventy five percent of the times, the correct user profile was in the top 3 results returned by our system. Our proposed user profile disambiguation system can help security analysts compare and analyze two different social networks. In future, we plan to incorporate more profile fields and generalize our model to make it applicable to include other social networks. We also want to adapt our system to handle missing and incorrect profile attributes. 

{\small{
\bibliographystyle{IEEEtran}
\bibliography{IEEEabrv,asonam}
}
%
\end{document}